\documentclass[%
 reprint,
amsmath,amssymb,
aps,
nofootinbib
]{revtex4-2}

\usepackage{graphicx}

\usepackage{braket}
\usepackage{bm}
\usepackage{amsmath}
\usepackage{algorithm}
\usepackage{algorithmic}
\usepackage{xcolor}
\usepackage{xspace}
\usepackage{subcaption}

\begin{document}

\preprint{APS/123-QED}

\title{Trainability maximization using estimation of distribution algorithms \\ assisted by surrogate modelling for quantum architecture search}

\author{Vicente P. Soloviev$^1$}
\email{(Corresponding author) vicente.perez.soloviev@fi.upm.es}
\author{Vedran Dunjko$^{2, 3}$}
\email{v.dunjko@liacs.leidenuniv.nl}
\author{Concha Bielza$^1$}
\email{mcbielza@fi.upm.es}
\author{Pedro Larrañaga$^1$}
\email{pedro.larranaga@fi.upm.es}
\author{Hao Wang$^{2, 3}$}
\email{h.wang@liacs.leidenuniv.nl}

\affiliation{$^{1}$Computational Intelligence Group (CIG), Universidad Politécnica de Madrid}
\affiliation{$^{2}$Applied Quantum Algorithms Leiden $\braket{aQa^L}$, Leiden University}
\affiliation{$^{3}$Leiden Institute of Advanced Computer Science (LIACS), Leiden University}

\begin{abstract}
Quantum architecture search (QAS) involves optimizing both the quantum parametric circuit configuration but also its parameters for a variational quantum algorithm. 
Thus, the problem is known to be multi-level as the performance of a given architecture is unknown until its parameters are tuned using classical routines. Moreover, the task becomes even more complicated since well-known trainability issues, e.g., barren plateaus (BPs), can occur.
In this paper, we aim to achieve two improvements in QAS: (1) to reduce the number of measurements by an online surrogate model of the evaluation process that aggressively discards architectures of poor performance; (2) to avoid training the circuits when BPs are present.
To detect the presence of the BPs, we employed a recently developed metric, information content, which only requires measuring the energy values of a small set of parameters to estimate the magnitude of cost function's gradient. The main idea of this proposal is to leverage a  recently developed metric which can be used to detect the onset of vanishing gradients to ensure the overall search avoids such unfavorable regions.
We experimentally validate our proposal for the variational quantum eigensolver and showcase that our algorithm is able to find solutions that have been previously proposed in the literature for the Hamiltonians; but also to outperform the state of the art when initializing the method from the set of architectures proposed in the literature. The results suggest that the proposed methodology could be used in environments where it is desired to improve the trainability of known architectures while maintaining good performance.
\end{abstract}

\keywords{Quantum architecture search, evolutionary algorithm, information content, Barren plateaus, estimation of distribution algorithm, multi-level optimization, variational quantum eigensolver}

\maketitle


\section{Introduction}
\label{sec_introduction}
Variational quantum algorithms (VQAs) \cite{bharti2022noisy} have become prominent tools in the noisy intermediate-scale quantum (NISQ) era, where quantum computers face limitations due to noise and connectivity issues. A well-known example of this type of approaches is the variational quantum eigensolver (VQE) \cite{peruzzo2014variational}.
Its adaptability and ability to efficiently explore solution spaces make them valuable tools for quantum computation, offering promising applications in areas such as quantum chemistry \cite{peruzzo2014variational}, optimization \cite{soloviev2022quantum}, and machine learning \cite{schuld2018supervised, wisniewska2023variational}, despite the challenges presented by the NISQ era hardware.

VQAs employ (i) an objective cost function to be minimized, (ii) a quantum parametric circuit (henceforth called as \textit{ansatz}), and (iii) a classical optimization technique that tunes the \textit{ansatz}.


First, a Hamiltonian ($H$) is a quantum Hermitian operator that describes a physical system, yielding the energy of a quantum state, which is often used as the objective cost function to be minimized in VQAs. Finding the global minima of the Hamiltonian (ground energy) implies finding a ground state of the quantum system. 
Although the literature proposes other objective functions such as the conditional value at a risk \cite{barkoutsos2020improving}, or the Gibbs objective function \cite{li2020quantum}, the most widely used one is the expectation value, often simplified as, 
\begin{equation}
\label{eq_expectation_value}
    \min_{\bm{\theta}} \quad \braket{H}_{U(\bm{\theta})},
\end{equation}
where $\bm{\theta}$ is the variational parameter, to be optimized classically, and $\braket{H}_{U(\bm{\theta})}$ describes the measurements of a quantum system as, 
\begin{equation}
    \braket{H}_{U(\bm{\theta})} = \bra{0} U^T(\bm{\theta}) H U(\bm{\theta}) \ket{0},
\end{equation}
where $U(\bm{\theta})$ is the unitary state generated by an \textit{ansatz}, parameterized by $\bm{\theta} \in [0, 2\pi]^d$, where $d$ is the number of parameters.


Second, an \textit{ansatz} is a quantum circuit which is parameterized by a set of parameters $\bm{\theta}$, and its quantum state is denoted as,
\begin{equation}
\label{eq_ansatz_unitary}
    \ket{\Psi(\bm{\theta})} = U(\bm{\theta}) \ket{\Psi_0},
\end{equation}
where $\ket{\Psi_0}$ is the given initial state, typically set to the $\ket{0}$ state, i.e., $\ket{00\cdots0}^{\otimes{n}}$ state, where $n$ is the number of qubits of the system.

The \textit{ansatz} found in the literature are traditionally classified into \textit{problem-inspired} or \textit{hardware-efficient}, depending on its design \cite{bharti2022noisy}. The former considers the intrinsic physics of the problem to be solved for its design, and it has been shown to achieve good performance in terms of quality and convergence. An example is the quantum approximate optimization algorithm \cite{farhi2014quantum}. However, the latter proposes \textit{ansatzes} that fit to the hardware limitations underlining a quantum device, i.e., available quantum gates or quantum connectivity. 


Third, the overall performance of the VQA heavily depends on both, \textit{ansatz} selection and the parameter optimization. Thus, the literature proposes a wide range of approaches to tune the parameters, which are typically classified into \textit{gradient-based} or \textit{gradient-free optimizers}. Some examples of the former include gradient descent \citep{ruder2016overview} and limited Broyden-Fletcher-Goldfarb-Shanno \citep{byrd1995limited}; while some examples of the latter include evolutionary algorithms (EAs) \cite{soloviev2023variational, spall1992multivariate} and reinforcement learning \cite{garcia2019quantum}, among others. 


When choosing an \textit{ansatz} for a problem and optimizing its parameters, we assume that the \textit{ansatz} is expressive enough to converge to the ground state of our Hamiltonian.
Finding the ideal \textit{ansatz} for a given $H$ but also the parameters $\bm{\theta}$ becomes a multi-level optimization problem \cite{mejia2023multiobjective} in which each proposed \textit{ansatz} also involves a new optimization task regarding the parameters of the specific architecture. 
Some approaches are presented in the literature using heuristics, where most of them involve too many measurements, and therefore lead to an increase of the computational resources and time. 
This is crucial for the feasibility of the algorithm in NISQ devices as the number of available measurements is limited before the device is re-configured.
Overcoming these limitations leads us to the quantum architecture search (QAS) research topic, where some authors have proposed different ideas. Further approaches regarding QAS are reviewed in Section~\ref{sec_related_work}.


The training/optimization of the variational parameters is known to be a non-trivial task for deep circuits, since we might face quite a few challenging trainability issues, e.g., BPs and traps \cite{anschuetz2022quantum}. 
BPs are typically described as vanishing gradients close to zero in the landscape, where the classical optimization becomes challenging, i.e., non-trainable or hard-to-train \textit{ansatz}. Several works are found in the state of the art where this phenomenon is studied in order to analyze the trainability of the \textit{ansatz} \cite{cerezo2021cost, mcclean2018barren}. 
However, computing these gradients involves the parameter optimization of the \textit{ansatz}, and thus increasing the number of quantum simulations, as we need to estimate the variance of the partial derivatives over the entire parameter space (exponential complexity). These tasks becomes more difficult with the number of qubits.
Recently, Pérez-Salinas et al. \cite{perez2023analyzing} have shown that the information content (IC) metric can reliably estimate the average (over the parameter space) norm of the gradient with a small number of evaluations of parameters of the \textit{ansatz}.



In this paper we propose a domain-agnostic approach based on EAs in which, given a set of \textit{ansatzes}, for which a good performance is expected, we seek to find a new set of \textit{ansatzes} similar to the initial one, but which are easier to train, and therefore are more likely to avoid the presence of BPs. The number of quantum simulations are drastically reduced by implementing a surrogate model which predicts the performance of the \textit{ansatz}, and the IC is used to maximize the trainability of the proposed architectures avoiding the presence of BPs. Experimental results are shown in noisy environments for different problems. Thus, the main contributions of the paper are:
\begin{itemize}
    \item The use of surrogate models to rank the \textit{ansatz} proposed by the EA without any measurements.
    \item The maximization of the trainability during the optimization process by using the IC.
    \item The use of multi-objective optimization to optimize the IC and the score provided by the surrogate model.
\end{itemize}
To the best of our knowledge this is the first work in which IC is optimized for quantum \textit{ansatz} design, and we conjecture this approach can pave the way to bridging the gap towards an ideal training-free approach.

The rest of the paper is organized as follows. Section~\ref{sec_related_work} reviews the QAS literature. In Section~\ref{sec_background} we provide a theoretical background for evolutionary approaches, IC for the approximation of the average norm of the gradients, and surrogate modelling. The proposed methodology is presented in Section~\ref{sec_method} and Section~\ref{sec_results} shows some experimental results. Section~\ref{sec_conclusions} rounds the paper off with some further conclusions and future open research lines. 

\section{Related work}
\label{sec_related_work}
This section reviews some of the existing works regarding QAS in the literature.

Regarding reinforcement learning (RL), \cite{pirhooshyaran2021quantum} uses a multi-level optimization process in which the agent proposes new architectures while a classical secondary optimizer tunes the parameters of the \textit{ansatz}. 
In \cite{fosel2021quantum}, a RL approach is proposed with a different purpose: given an \textit{ansatz}, return an optimized structure in terms of circuit depth and used gates. 
A RL approach is proposed \cite{ostaszewski2021reinforcement} where an agent systematically modifies the \textit{ansatz} and achieves shallow circuits for chemical domains. More recently, a novel approach based in RL is proposed in \cite{patel2024curriculum} with competitive results.

Regarding EAs, 
\cite{chivilikhin2020mog} proposes a multi-level genetic algorithm where a multi-objective approach is used to minimize the energy of the VQE while minimizing the number of CNOT gates, and the parameter optimization is performed by CMA-ES optimizer.
In \cite{rattew2019domain} the authors use a genetic algorithm to optimize a weighted single-objective cost function combining the energy of the proposed \textit{ansatz}, its depth, and number of two-qubit gates. 
Recently, GA4QCO framework \cite{sunkel2023ga4qco} is proposed in which a single-objective optimization is performed by a genetic algorithm, and compared to random instances.

Regarding chemistry simulation, AdaptiveVQE \cite{grimsley2019adaptive} is a methodology that systematically grows an \textit{ansatz} for chemical simulation; and RotoSelect and RotoSolve methods \cite{ostaszewski2021structure} are two efficient methods for jointly optimizing \textit{ansatz} structure and parameters.

Several works are found in the literature in which neural architecture search methodologies are applied to QAS. 
QuantumDARTS \cite{wu2023quantumdarts} is an adaptation of classical DARTS \cite{liu2018darts} for neural network architecture search to QAS, in which two methods are proposed: one for whole architecture search, and another for promising sub-architectures. 
Another example is \cite{zhang2022differentiable} in which new architectures are sampled from a probabilistic model, and gradients between the best energies found are computed. 


Additionally, SuperNet structure \cite{du2020quantum}, samples several architectures and its parameters are classically optimized. Based on the performance, the \textit{ansatz} are ranked and a new architecture is constructed based on the knowledge gained from them. SuperNet has also been used to enhance VQAs on an 8-qubit superconducting quantum processor for classification tasks \cite{linghu2022quantum}. 

Our work is an EA which differs from the rest by using a multi-objective approach, reducing the complexity of the multi-level optimization task by using surrogate modeling and information content to evaluate the presence of BPs.

\section{Background}
\label{sec_background}
\subsection{Estimation of distribution algorithms}
EAs are a class of optimization and search techniques inspired by the principles of natural selection and biological evolution. Rooted in the idea of survival of the fittest, these algorithms mimic the process of evolution to iteratively improve and evolve a population of candidate solutions to a problem. Traditional EAs rely on crossover and mutation operators, whereas, estimation of distribution algorithms (EDAs) \cite{larranaga2001estimation} iteratively learn and sample unclear modelling what target probability distribution. EDAs have shown to be a power tool for optimization problems in which the number of variables to be optimized is big.

\begin{algorithm}[H]
    \caption{Estimation of distribution algorithms}
    \label{alg_eda_baseline}
    \textbf{Input}: Population size $N$, selection ratio $\alpha$, cost function $g$ \\
    \textbf{Output}: Best individual $\bm{x}'$ and cost found $g(\bm{x}')$ \\
    
    \begin{algorithmic}[1]
        \STATE $G_0 \leftarrow $ $N$ individuals randomly sampled or provided
        \FOR{$t = 1,2, ... $ until stopping criterion is met}
            \STATE Evaluate $G_{t-1}$ according to $g(\cdot)$
            \STATE $G_{t-1}^{S} \leftarrow $ Select top $\lfloor \alpha N \rfloor$ individuals from $G_{t-1}$
            \STATE $p_{t-1} \leftarrow $ Learn a probabilistic model from $G_{t-1}^{S}$
            \STATE $G_t \leftarrow $ Sample $N$ individuals from $p_{t-1}(\cdot)$
        \ENDFOR
    \end{algorithmic}
\end{algorithm}

Algorithm~\ref{alg_eda_baseline} describes the baseline of EDA approaches.
Given a population of size $N$, the ratio of the population $\alpha \in (0,1)$ to be promoted to next iteration, and the cost function $g(\cdot)$ to be optimized, the algorithm iteratively selects the top $\lfloor \alpha N \rfloor$ individuals from a set of solutions according to $g(\cdot)$ (lines 3-4), learns a probabilistic model (line 5) from these top individuals, and samples it to generate a new set of solutions (line 6).
The algorithm iterates until a convergence criterion is met, and returns the best cost and solution found so far.

Regarding the type of probabilistic model, we can distinguish between \textit{multivariate} EDAs and \textit{univariate} EDAs. The former learns a joint probability distribution factorized with conditional probabilities over the variables involved in the problem. The latter learns a univariate probability distribution per variable in which no dependencies are considered, speeding up the computation and thus allowing to face bigger optimization problems, in terms of the number of variables.

Considering the set of random variables $\bm{X} = (X_1, X_2, \dots, X_d)$ involved in the problem, where $d$ regards the dimension of the feature space, the joint probability distribution is approximated in the univariate EDAs as,
\begin{equation}
\label{eq_joint_probability_univariate}
    p(\bm{X}) = p(X_1, X_2, \dots, X_d) = \prod_{i=1}^d p(X_i),
\end{equation}
where $p(X_i)$ is the marginal probability distribution of variable $X_i$. Note that computing the joint probability distribution of multivariate EDAs is much more costly, and thus in this approach we use univariate EDAs.

\subsection{Information content for BPs diagnosis}
\label{sec_info_content_bp}
BPs are traditionally described as exponentially vanishing gradients of the cost function where a classical optimizer is placed in a flat landscape, in which finding the global optimum becomes challenging. Avoiding this type of landscapes increases the probability of reaching better solutions. However, computing the gradients involves optimizing the \textit{ansatz}, and thus, drastically increasing the number of quantum simulations.

Formally, BPs are characterized by the following properties,
\begin{equation}
    \mathbb{E}_{\theta} (\partial_k E(\bm{\theta})) = 0,
\end{equation}
\begin{equation}
    \text{Var}(\partial_k E(\bm{\theta})) \in \mathcal{O}(\text{exp}(-n)) , 
\end{equation}
where $\mathbb{E} (\partial_k E(\bm{\theta}))$, $k \in [1 \dots m]$, and $\operatorname{Var}(\partial_k E(\bm{\theta}))$ are the expectation and variance of the partial derivatives of the objective cost function, respectively, $\bm{\theta}$ is the set of parameters of the unitary representing the \textit{ansatz}, and $n$ is the number of qubits. 

Recently, Pérez-Salinas et al. \cite{perez2023analyzing} have shown that the norm of the gradients can be bounded efficiently with a small number of quantum measurements  (which grows linearly with the number of parameters), without the need of optimizing the \textit{ansatz} parameters. This method performs a random walk in the parameter space and measures the entropy of fluctuations of cost values along the walk. The measured entropy value can be used to analytically bound the gradient of the cost function along the walk. We notice that the average of the gradient field (henceforth named as IC) can be approximated by the average along the random walk (due to Monte Carlo integration):
\begin{equation}
\label{eq_information_content}
    \parallel \nabla E \parallel ^2 \approx \mathbb{E}_W \left( \sum^m_{k=1} (\partial_k E(\bm{\theta}))^2 \right) = \sum^m_{k=1} \text{Var}_W(\partial_k E(\bm{\theta})),
\end{equation}
where $\text{Var}_W$ denotes the variance found in the objective cost function using $m$ different $\bm{\theta}$ parameters generated from a random walk $W$. Note that this sampling is more efficient than estimating the gradients from random points.

Therefore, we propose to measure the IC metric for each candidate architecture, and maximize the IC value across the architecture search in addition to minimizing the cost value. This approach can help the architecture to generate more trainable circuits. 

\begin{figure*}[t]
    \centering
    \includegraphics[width=0.85\linewidth]{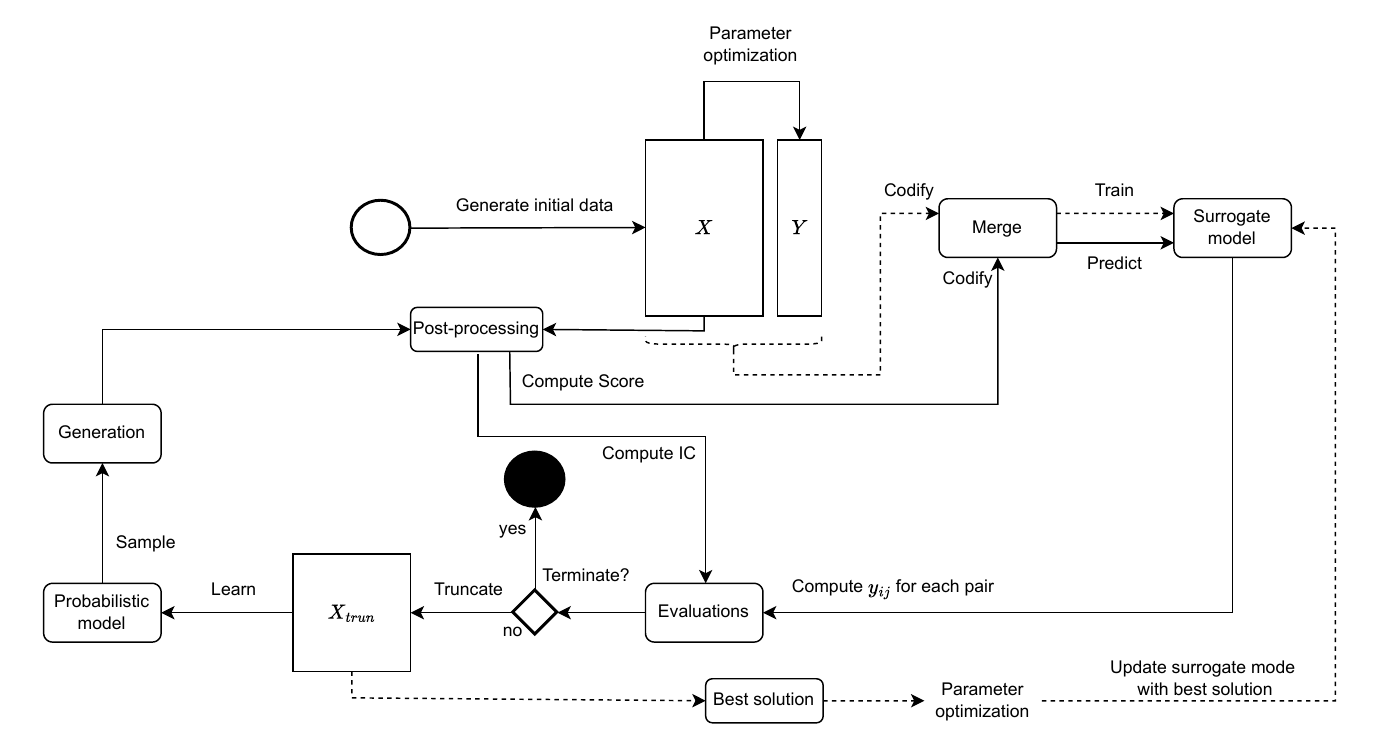}
    \caption{Flowchart of the proposed approach, starting from the white spot and finishing in the black spot one the convergence criteria is met. Dashed lines regard the train and update of the surrogate model.}
    \label{fig_flow_chat_mo_eda_qas}
\end{figure*}

\subsection{Surrogate modelling}
Surrogate modelling is a common approach in machine learning for approximating the performance of an expensive computational task. Formally, we define a surrogate model as a function $h'(\bm{X})$ that approximates the output of $h(\bm{X})$, where $\bm{X} = (X_1, X_2, \dots, X_d)$ is the input space with dimension $d$, and $h(\cdot)$ is a multivariate function that is time consuming to compute. The surrogate model $h'(\cdot)$ is formulated to provide a computationally efficient alternative and as a supervised approach it is constructed based on a set of observed data points $\mathcal{D} = \{( \bm{x}_i, h(\bm{x}_i)) \}^S_{i=1}$, where $\bm{x}_i$ is an instance of the dataset with associated performance $h(\bm{x}_i)$, and $S$ is the number of instances in the dataset.


\section{Method}
\label{sec_method}
This section explains the proposed approach and describes each of the modules in the following subsections. Figure~\ref{fig_flow_chat_mo_eda_qas} summarizes the flowchart of the approach where the main steps of the proposed algorithm are stated.

\subsection{Codification}
For an \textit{ansatz} of $n$ qubits and maximally depth $m$,  we propose the following integer-valued matrix representation:
\begin{align}
\label{eq_codification}
    \bm{X} &= 
    \begin{bmatrix}
    X_{11} & \cdots & X_{1m} \\
    \vdots & \ddots & \vdots \\
    X_{n1} & \cdots & X_{nm}
    \end{bmatrix} \\ \nonumber
    &\rightarrow [X_{11}, \cdots, X_{1m}, \cdots, X_{n1}, \cdots, X_{nm}],
\end{align}
where each entry $X_{ij} \in \{0, 1, \dots, n_{gates}\}$ represents the choice of the quantum logic gate at position $(i, j)$ of the matrix. Given a predetermined number of qubits $n$ and maximal depth $m$, the architecture representation has a fixed dimension $d=nm$. This way, each column represents all the operators executed in parallel along the total depth, and each row represents a qubit.

Note that regarding two-qubit gates such as CNOT, applying a CNOT with the same control qubit, but different target qubits, are considered as different gates. This allows to restrict the evolutionary search according to hardware constraints by restricting the search space, although in this work an all-to-all connectivity is considered. In our case, $n_{gates} = (n-1) + 5$, as we consider the following universal operators: $\{Rx(\cdot), Ry(\cdot), Rz(\cdot), H, I\}$ and the CNOT gate with different target qubits. Note that CNOT$(i, j)$ denotes that $i$ and $j$ are the control and target qubits, respectively.

The initial state of all the proposed architectures is set to the $\ket{0}$ state, i.e., $\ket{00\cdots0}^{\otimes{n}}$ state.

Figure~\ref{fig_preprocess} shows an example where the following codification is represented as an \textit{ansatz},
\begin{equation}
\label{eq_example_ansatz}
    \bm{A} = 
    \begin{bmatrix}
        4 & 0 & 1 & 3 \\
        4 & 4 & 5 & 2 \\
        2 & 2 & 5 & 5
    \end{bmatrix},
\end{equation}
where $n=3$ and $m=4$.


\subsection{Probabilistic model}
The joint probability distribution factorizes in a univariate EDA approach according to Equation~\ref{eq_joint_probability_univariate}, where $p(X_{ij})$ is the marginal probability distribution of variable $X_{ij}$. In this approach, $d=nm$, and $p(X_{ij})$ follows a multinomial distribution,
\begin{equation}
    X_{ij} \sim \text{Mult}(n_m=\lfloor \alpha N \rfloor, k_m=(n_{gates}+1)),
\end{equation}
where $n_m$ and $k_m$ are the number of trials and mutually exclusive events that define the multinomial probability distribution, respectively.

Note that the marginal probabilities over the set of solutions are computed after the truncation process (Algorithm~\ref{alg_eda_baseline} Line~4), where the top $\lfloor \alpha N \rfloor$ solutions are selected according to the cost function to be optimized. The sampling process generates $N$ new solutions as detailed in Algorithm~\ref{alg_eda_baseline}, and duplicate \textit{ansatz} are rejected in order to reduce redundancy. Each solution represents an \textit{ansatz}, and the algorithm is expected to learn itself the best gates configuration during runtime. 

\subsection{Post-processing}
\label{sec_post_processing}
In order to restrict the search space of the QAS problem, we establish a series of hard rules to remove redundancy and simplify the \textit{ansatz} architectures proposed in the sampling process of the EDA. 
\begin{itemize}
    \item Two consecutive $H$ gates are removed, as they are equivalent to an $I$ gate.
    \item Consecutive application of $Rx(\cdot)$ gates, are simplified as one single $Rx(\cdot)$ gate, to remove redundancy. 
    \item Consecutive application of $Ry(\cdot)$ gates, are simplified as one single $Ry(\cdot)$ gate, to remove redundancy. 
    \item Consecutive application of $Rz(\cdot)$ gates, are simplified as one single $Rz(\cdot)$ gate, to remove redundancy. 
\end{itemize}

\begin{figure}[t]
    \centering
    \begin{subfigure}{0.23\textwidth}
        \includegraphics[width=\textwidth]{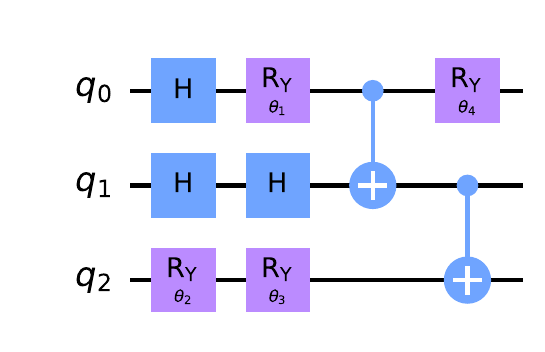}
        \caption{Original \textit{ansatz}}
        \label{fig_preprocess}
    \end{subfigure}
    \hfill
    \begin{subfigure}{0.23\textwidth}
        \includegraphics[width=\textwidth]{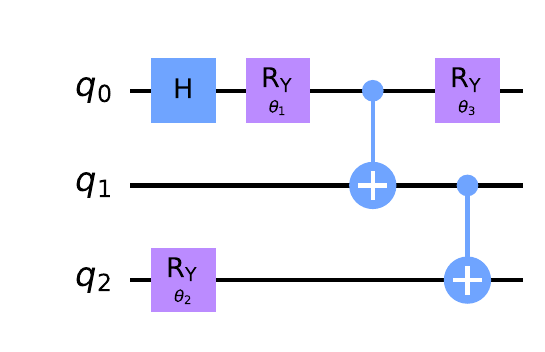}
        \caption{Post-processed \textit{ansatz}}
        \label{fig_postprocess}
    \end{subfigure}

    \caption{Post-processing of an \textit{ansatz} where hard rules (Section~\ref{sec_post_processing}) have been applied to the architecture represented in Equation~\ref{eq_example_ansatz} with $n=3$ and $m=4$.}
    \label{fig_cod_process_qas}
\end{figure}

Once the algorithm samples a new set of architectures (Algorithm~\ref{alg_eda_baseline} Line~6), the post-processing step is applied to each of them. Figure~\ref{fig_cod_process_qas} shows an example of the application of these hard rules, where (i) in the second qubit, both consecutive $H$ gates were suppressed, and (ii) in the third qubit the two $Ry(\cdot)$ gates are simplified as a single gate.

\subsection{Surrogate model}
\label{sec_surrogate_model_eda}
A characteristic of traditional EDAs is that once the solutions of the same population are ranked according to $g(\cdot)$, no matter how much better a solution is compared to others, as all solutions included in the top $\lfloor \alpha N \rfloor$ will contribute equally to the probabilistic model learning \cite{larranaga2023estimation} (see Algorithm~\ref{alg_eda_baseline}, Line~4).
The surrogate model used in this approach surrogates the minimal thing needed for the EDA, that is, the ranking of solutions (line 4 Algorithm~\ref{alg_eda_baseline}). This is introduced by a metric $\text{Score}(A)$ (inspired in \cite{shi2021fast}) which measures the quality of a solution $A$ within the rest of solutions of the population,
\begin{equation}
\label{eq_score_a}
    \text{Score}(A) = \sum_{B \in \bm{X}} ( h(A, B) + 1 - h(B, A) ),
\end{equation}
where the higher $\text{Score}(A)$, the better the quality of $A$, and $h(A, B)$ compares \textit{ansatz} $A$ to \textit{ansatz} $B$ as,
\begin{equation}
    h(A, B) = 
    \begin{cases}
    0, & \text{if \;\;} P_B \geq P_A + \epsilon \\
    1, & \text{if \;\;} P_A \geq P_B + \epsilon \\
    2, & \text{otherwise}
    \end{cases}
\end{equation}
where $P_A$, and $P_B$ are the minimum expectation values (Equation~\ref{eq_expectation_value}) found by a classical optimizer for architectures $A$ and $B$, respectively and $\epsilon$ is a tolerance error configured by the user. Note that $h(A, B) = h(B, A) = 2$ means that two \textit{ansatz} $A$ and $B$ are non comparable or very similar performance is expected. 

Computing $\text{Score}(A)$ involves $\lfloor \alpha N \rfloor - 1$ comparisons, and thus, this is clearly the main bottleneck of the task. In order to overcome this, we propose the use of support vector machines (SVMs) to approximate $h(A, B)$. We take the following input feature to the surrogate model: 
\begin{align}
    \operatorname{Flatten}(A + B, A - B)
\end{align}
where $A$ and $B$ are the two \textit{ansatz} architectures to be compared, and the resultant vector size is $d=2nm$. Thus, $h(A, B) \in \{0, 1, 2\}$ is approximated by $h'(\text{Flatten}(A, B)) \in \{0, 1, 2\}$ using SVM.


Several classification methods have been tested over some initial data randomly generated for different values of $n$, where SVM achieved better accuracy metrics. Results using cross-validation can be found in Appendix~\ref{sec_appendix_surrogate_model_prediction}.

The implementation has been obtained from LibSVM library \cite{chang2011libsvm}.

The surrogate model is re-fitted after each iteration with the top 5 solutions in the ranking of the best solutions computed by the EDA (Section~\ref{sec_evaluation_eda}). Thus, in each iteration 5 classical parameter optimizations are carried out, and the number of parameter tuning processes executed during runtime is $N + 5t$, where $t$ is the total number of iterations. Without the usage of the surrogate model approach, this number would have been $N(1 + t)$.


\subsection{Evaluation}
\label{sec_evaluation_eda}
This approach aims to find the optimal \textit{ansatz} for a given problem $H$ in terms of trainability and expected energy. Here we define the following metrics to be computed for each proposed architecture.

First, IC (Equation~\ref{eq_information_content}) maximization has been proved to be able to avoid BP in the \textit{ansatz} parameter tuning \cite{perez2023analyzing}. Those architectures with low associated IC are less trainable/optimizable, compared to those with high IC. Our approach maximizes this metric through the optimization process. 
Here, the IC of an ansatz $A$ is denoted as,
\begin{equation}
    \text{IC}(A) = \epsilon_M \sqrt{M},
\end{equation}
where $\epsilon_M$ is the $\epsilon$ associated to the norm of the gradient computed after a random walk over the parameters (Section~\ref{sec_info_content_bp}), 
and $M$ is the number of parameters of \textit{ansatz} $A$.

Second, $\text{Score}(\cdot)$ (Equation~\ref{eq_score_a}) evaluates the quality of a solution compared to a subset of solutions. Our approach implements an elite approach, in which the best solution of generation $G_i$ also appears in generation $G_{i+1}$. Then finding a different best solution in $G_{i+1}$ will lead to a best global solution in the whole optimization process. Thus, $\text{Score}(\cdot)$ is also desired to be maximized. 

Maximizing both metrics becomes a multi-objective optimization problem, in which the Pareto frontier between both objectives is explored. During the optimization process defined in Algorithm~\ref{alg_eda_baseline} and Figure~\ref{fig_flow_chat_mo_eda_qas}, the truncation process ranks the solutions according to $g(\cdot)$, which is here defined as,
\begin{equation}
    \label{eq_final_evaluation}
    g(A) = \text{HV}((\text{Score}(A), \text{IC}(A)), \; \bm{r}),
\end{equation}
where $\text{HV}(\cdot)$ is the hypervolume contribution \cite{beume2009complexity} between the surrogate model output ($\text{Score}(A))$ and the information content computed ($\text{IC}(A)$), and $\bm{r}$ is the reference point. The $\lfloor \alpha N \rfloor$ best solutions in terms of $\text{HV}(\cdot)$ minimization are the ones that better approximate the Pareto frontier, and are the ones that promote to the next EDA iteration.

The reference point can be estimated based on the bounds of $\text{Score}(A)$ and $\text{IC}(A)$. In the former, the lower bound is set to zero (the worst solution within the population) and the upper bound to $2N$ (the best solution within the population). In the latter, the lower bound is set to zero (the least trainable scenario) and the upper bound to $2$, based on previous experience. Then, $\text{Score}(A) \in \{0, 1, \dots, 2N\}$ and $\text{IC}(A) \in [0, 2] \in \mathbb{R}$, so the reference point is set to $\bm{r} = (2N, 2)$.

Finally, the optimization problem is formalized as,
\begin{equation}
\begin{split}
    &\min_{\bm{X}} \quad g(\bm{X}) \\
    &\text{subject to } \bm{X} \in \{0, 1, \dots, n_{gates}\},
\end{split}
\end{equation}
where $\bm{X}$ denotes a codified \textit{ansatz} (Equation~\ref{eq_codification}), and $g(\dot)$ is defined at Equation~\ref{eq_final_evaluation}. 

\section{Results}
\label{sec_results}
This section shows some numerical results on solving different Hamiltonians $H \in \{H_1, H_2, H_3, H_4\}$ (Appendix~\ref{sec_appendix_hamiltonians}), already studied in \cite{nakayama2023vqegenerated} for $n \in \{4, 8, 12\}$. The following sections compare the results found by the EDA approach with those presented in the dataset from \cite{nakayama2023vqegenerated}. In the original paper, the authors present several architectures which find similar state vectors in the search space of VQE \textit{ansatz}, for each $H_i$. Henceforth, $D_i^n$ denotes the set of architectures proposed in the dataset to solve the Hamiltonian $H_i$ with $n$ qubits. 

Two experiments have been carried out in which, (i) the initial population of the EDA approach is initialized randomly to test if the algorithm is able to converge to similar solutions to those proposed in the dataset (Section~\ref{sec_random_init}), and (ii) the initial population is initialized from the \textit{ansatzes} proposed in the dataset \cite{nakayama2023vqegenerated} to test if the algorithm is able to improve the given architectures (Section~\ref{sec_data_init}).

The size of the population, and maximum number of iterations of the EDA have been set to $N=150$ and $t=50$, respectively, for all the experiments. Regarding the quantum circuit simulation, we simulate the measurement noise.

\begin{figure}[t]
    \centering
    \includegraphics[width=1\linewidth]{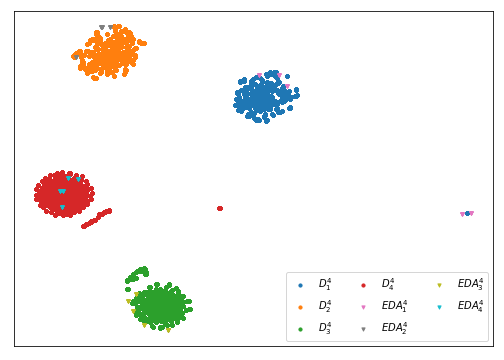}
    \caption{Visualization of the \textit{ansatzes} found in the dataset ($D_i^n$) using t-SNE \cite{van2008visualizing} , which are colored depending on the Hamiltonian to be solved ($H_i$, where $i \in \{1, 2, 3, 4\}$). Additionally, the best architectures found by the EDA approach (EDA$^n_i$) are represented using different colored and shaped points. Note that EDA$^n_i$ regards the solutions found for Hamiltonian $H_i$. All the results shown correspond to $n=4$.}
    \label{fig_clusters_state_vectors}
\end{figure}

\subsection{Random initialization}
\label{sec_random_init}
To randomly generate the initial population ($G_0$), a predefined probabilistic model is set to the algorithm, from which the set of solutions are sampled. Thus, some of the outcomes for each variable can be restricted, or boosted, decreasing or increasing the associated probabilities, respectively, as demanded by the user.  

In this experiment, initially, all the possible outcomes have been set to equal probability for all the variables:
\begin{equation}
\label{eq_equal_probabilities_initial_pop}
    p(X_i = j) = \frac{1}{n_{gates} + 1},
\end{equation}
for all $i=1, \dots, d$ and $j=0, 1, \dots, n_{gates}$.

\begin{figure*}[t]
    \centering
    \begin{subfigure}{0.32\textwidth}
        \includegraphics[width=\textwidth]{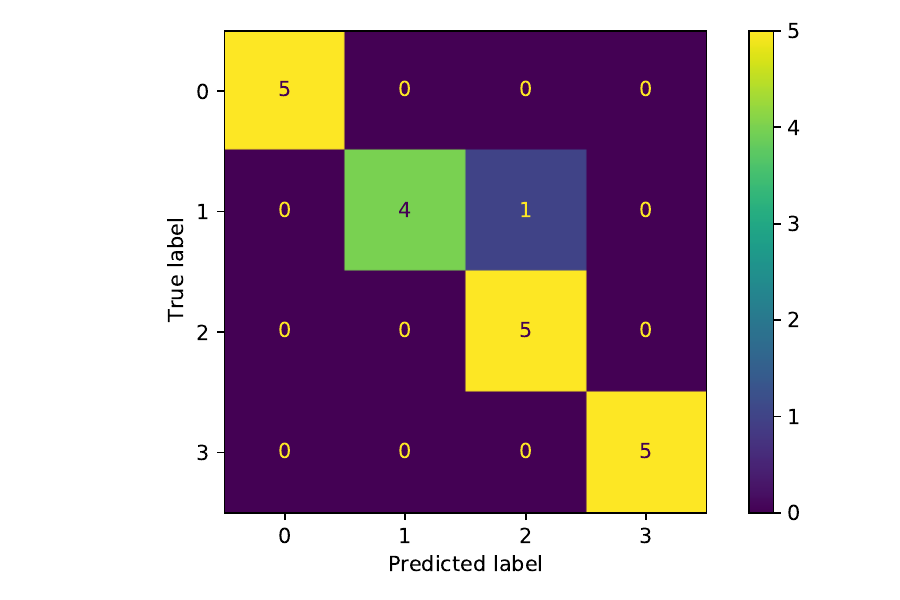}
        \caption{$n=4$}
        \label{fig_confusion_matrix_4}
    \end{subfigure}
    \hfill
    \begin{subfigure}{0.32\textwidth}
        \includegraphics[width=\textwidth]{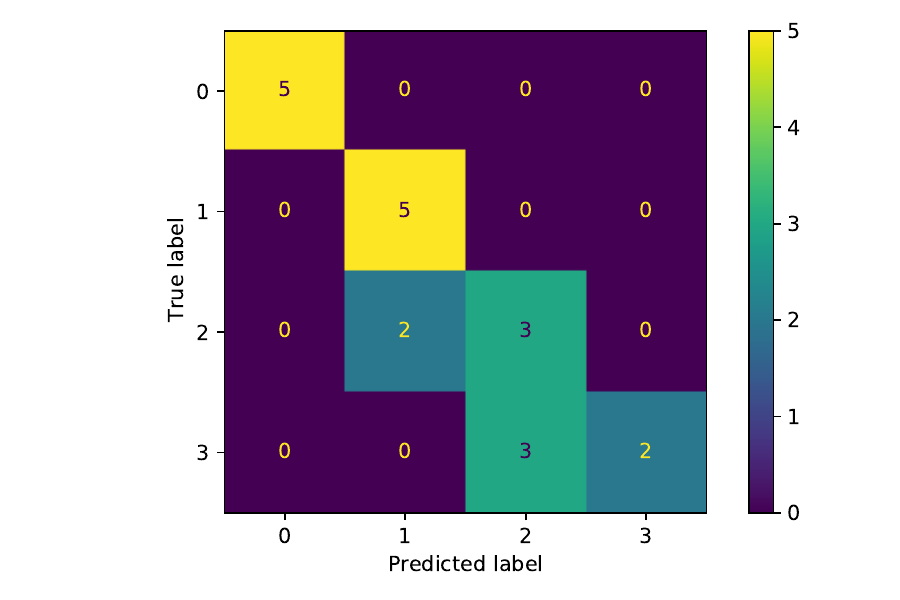}
        \caption{$n=8$}
        \label{fig_confusion_matrix_8}
    \end{subfigure}
    \hfill\begin{subfigure}{0.32\textwidth}
        \includegraphics[width=\textwidth]{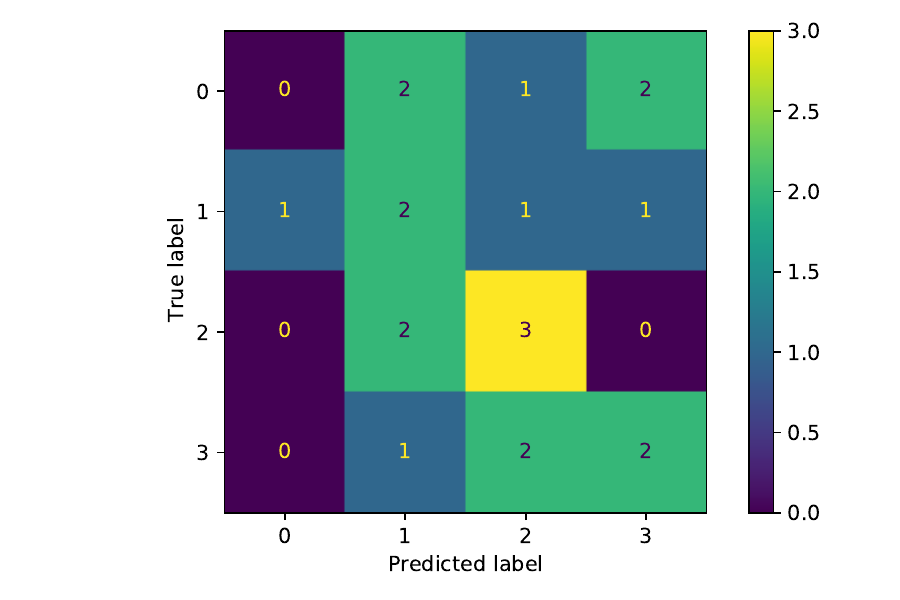}
        \caption{$n=12$}
        \label{fig_confusion_matrix_12}
    \end{subfigure}
    \caption{Confusion matrices for $n \in \{4, 8, 12\}$.}
    \label{fig_confusion_matrices}
\end{figure*}

The initial population samples a set of $N$ solutions, according to Equation~\ref{eq_equal_probabilities_initial_pop}. Each sample corresponds to a different architecture following the codification in Equation~\ref{eq_codification} and is post-processed (Section~\ref{sec_post_processing}). The expectation value (Equation~\ref{eq_expectation_value}) of each architecture is computed, where its parameters are classically optimized using an external optimizer. In this experiment we use COBYLA optimizer, as it has been shown to achieve good results in terms of CPU time and energy minimization \cite{powell1998direct}. Considering the set of solutions and associated expectation values, a surrogate model is trained (Section~\ref{sec_surrogate_model_eda}) and each solution is evaluated (Section~\ref{sec_evaluation_eda}).

The original dataset \citep{nakayama2023vqegenerated} proposes using dimensionality reduction to demonstrate that the minimal energy states achieved within $D_i^n$ are very similar. Figure~\ref{fig_clusters_state_vectors} shows the dimensional reduction using t-SNE \cite{van2008visualizing} for the Hamiltonians approached, represented as clusters in two dimensions. The solutions found by the EDA approach ($\text{EDA}_i^n$, where $i$ denotes the index of the faced Hamiltonian and $n$ the number of qubits) are also represented by stars and different colors. Note that our approach is able to reach very similar solutions to the ones presented in the dataset. 

In the following analysis the fidelity of the lowest energy state found by the EDA approach is compared to those obtained by the \textit{ansatzes} provided in the dataset for different problems $\{H_1, H_2, H_3, H_4\}$ and number of qubits ($n$), that is, by $D_i^n$. 

The distance from each proposed \textit{ansatz} ($A$) in $\text{EDA}_i^n$ to each cluster of architectures $D_i^n$ is computed by the arithmetic mean distance to each of the \textit{ansatzes} belonging to $D_i^n$ as,
\begin{equation}
\label{eq_distance_accuracy}
    \text{dist}(A, D_i^n) = \frac{1}{|D_i^n|}( \sum_{B \in D_i^n} 1 - F(\ket{\Psi_A}, \ket{\Psi_B}) ),
\end{equation}
where $D_i^n$ is the subset of \textit{ansatzes} (with size $|D_i|$) in the dataset proposed to solve $H_i$ with $n$ qubits and meet $m \pm \sqrt{m}$ restriction, $F(\cdot)$ is the fidelity between two quantum states, and $\ket{\Psi_A}$ and $\ket{\Psi_B}$ are the lowest energy states achieved by \textit{ansatzes} $A$ and $B$, respectively, after classical parameter optimization.

\begin{table}[h]
    \centering
    \setlength{\tabcolsep}{15pt}
    \begin{tabular}{|c|c|c|c|}
    \hline
        $H_i$ & $n=4$ & $n=8$ & $n=12$ \\
    \hline
    \hline
        $H_1$ & \textbf{3.0e-34} & \textbf{3.0e-2} & 6.0e-1 \\
        $H_2$ & \textbf{1.3e-4}  & \textbf{1.1e-2} & 1.5e-1 \\
        $H_3$ & \textbf{1.0e-15} & 3.0e-1 & 1.1e-1 \\
        $H_4$ & \textbf{2.0e-8}  & \textbf{5.1e-2} & 2.1e-1 \\
    \hline
    \end{tabular}
    \caption{ANOVA one-way test to reject the null hypothesis of equal means between the mean distances (Equation~\ref{eq_distance_accuracy}), from the proposed by \cite{nakayama2023vqegenerated} \textit{ansatzes} found by EDAs and $\{ D_1^n, D_2^n, D_3^n, D_4^n \}$ proposed for $\{H_1, H_2, H_3, H_4\}$, respectively. A threshold of 5e-2 has been set to reject the null hypothesis, highlighting in bold those results below this value.}
    \label{tab_anova_tests}
\end{table}

Table~\ref{tab_anova_tests} shows the $p$-values computed using the ANOVA test\footnote{All the data used for the ANOVA tests fit Gaussian distributions.} to reject the null hypothesis of equal means between each \textit{ansatz} in $\text{EDA}_i^n$ and the different clusters $D_i^n$, where highlighted results are rejected. Appendix~\ref{sec_appendix_distances} details the distance computations statistically analyzed in this table. An increasing number of non-rejected hypotheses is observed for increasing number of qubits ($n$), which suggests that the EDA is proposing architectures much different to the ones available at the dataset for $n=12$. Increasing the number of qubits ($n$) also involves increasing the number of variables of the EDA optimizer. According to the results found, the population size set is not enough to generate a large number of samples which covers the increasing cardinality of the problem. Also, larger number of qubits should also involve a larger \textit{ansatz} depth, so $m$ should also be increased to allow more expressive quantum circuits. This suggests that the chosen configuration is valid to problems up to $n<8$. For bigger instances, a different configuration of the hyper-parameters $m$ and $N$ should be chosen, although this would involve a drastic increase of the CPU time.

Assuming that a truly classified \textit{ansatz} ($A$) is the case in which the closest cluster $D_i^n$ represents $H_i$, and $A \in \text{EDA}_i^n$ was optimized for Hamiltonian $H_i$ as well, Figure~\ref{fig_confusion_matrices} shows the confusion matrices. The percentage of correctly classified \textit{ansatzes} is 95\%, 75\% and 35\% for $n = 4, 8, 12$, respectively, where a decreasing tendency is observed for increasing $n$;
however, for $n=12$ the EDA was not able to found any statistical significant result. 

\begin{figure}[t]
    \centering
    \includegraphics[width=\linewidth]{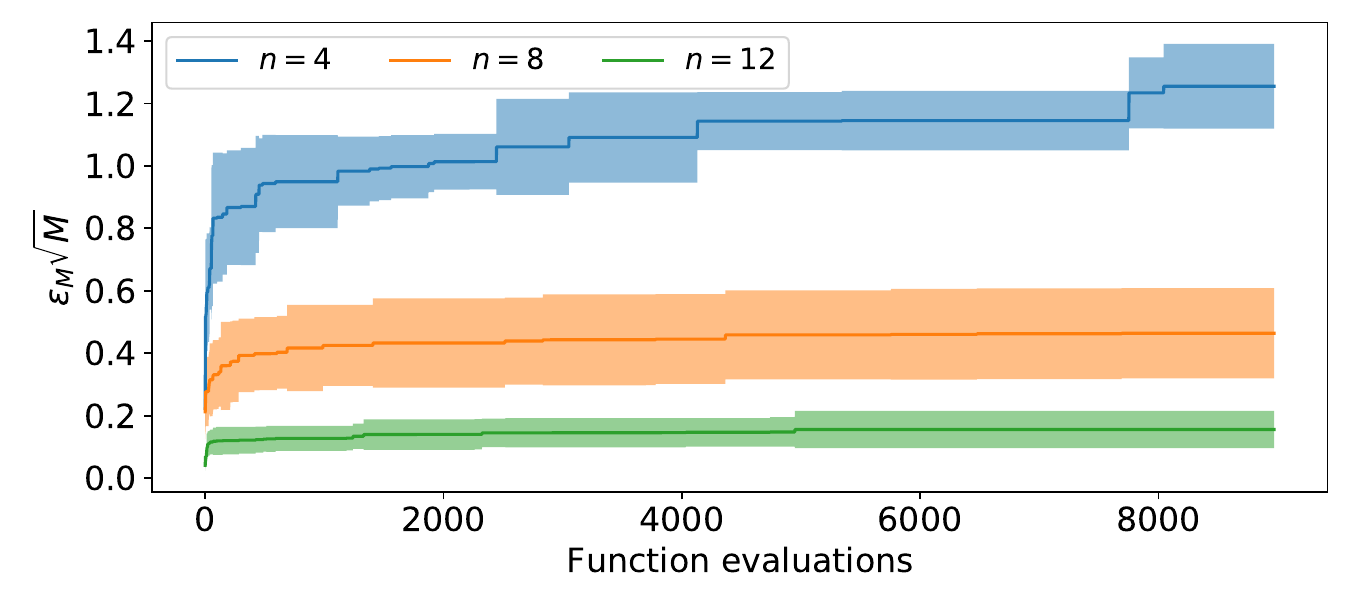}
    \caption{Mean and standard deviation of IC maximization aggregating the optimization process of different $H_i$ for different numbers of qubits ($n$).}
    \label{fig_IC_convergence_fevals_general}
\end{figure}

Figure~\ref{fig_IC_convergence_fevals_general} shows the IC convergence plot during the optimization process of the EDA approach. The associated shade shows a mean aggregation of the optimization processes regarding different $\{H_1, H_2, H_3, H_4\}$, where a maximizing monotonic tendency is observed. Regardless of the results encountered, the three scenarios show that the algorithm has converged. Note that, the mean IC found by the optimizer denotes an exponential decay with the number of qubits ($n$), as expected according to \cite{cerezo2021cost, perez2023analyzing}. 

Because $\text{Score}(A)$ returns a metric comparing \textit{ansatz} $A$ with the rest of the architectures within the population to which $A$ belongs, the trend throughout the optimization process is not an interesting fact to analyze. 

Appendix~\ref{sec_appendix_pareto_frontier_qubits_label} shows the Pareto frontier approximation (non-dominated solutions highlighted as orange spots) for each $H_i$ we are facing (in columns) and different values of $n$ (in rows). It is observed how both objectives are conflicting, and maximizing one of the objectives worsens the second, and vice-versa. Thus, a trade-off between both objectives through the Pareto frontier approximation is desired. Note that the scale of the Y-axis (IC) is different for different number of qubits, as explained before.

Considering the best solutions found by the EDA, i.e., those that better approximate the Pareto frontier, we now compare the characteristics of the \textit{ansatzes} proposals with those available in the dataset \cite{nakayama2023vqegenerated} with depth in the range $m \pm \sqrt{m}$ (for a fair comparison and ensure a minimum number of instances from the original dataset). A drastic increase in the number of certain quantum gates might improve the performance of the \textit{ansatz}, however, this may lead to a poor trainability. Thus, the ratio among the gates set used, and the number of gates is further analyzed.

\begin{figure}[t]
    \centering
    \includegraphics[width=\linewidth]{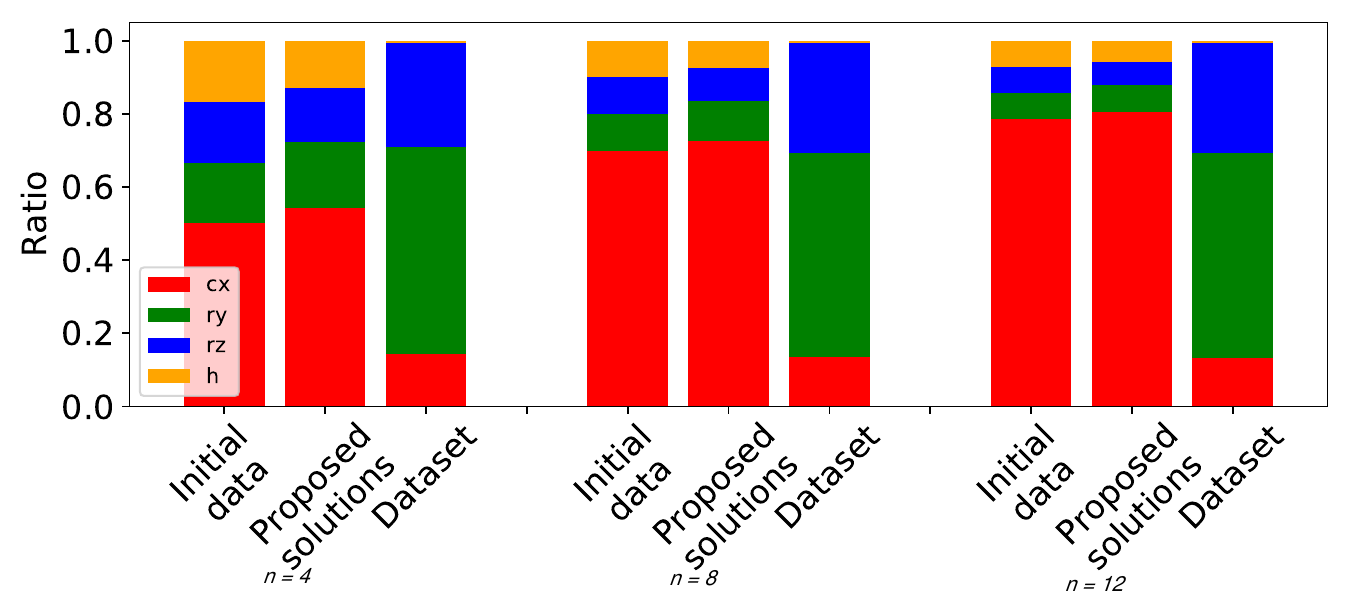}
    \caption{Ratio of $\{ CNOT, RY, RZ, H \}$ gates in the \textit{ansatz} design of the randomly generated initial data ($G_0$), best EDA solutions found ($\text{EDA}_i^n$), and dataset ($D_i^n$) \cite{nakayama2023vqegenerated}, for $n \in \{4, 8, 12\}$, respectively.}
    \label{fig_ratio_number_gates}
\end{figure}

Figure~\ref{fig_ratio_number_gates} shows the ratio of the different available universal gates in the set of initial randomly generated data ($G_0$), the solutions found by EDA approach ($\text{EDA}_i^n$) and the best solutions from the original dataset ($D_i^n$), for different values of $n$. A strong correlation is observed between the initial data and the proposed solutions, independently of $n$, where the $\text{EDA}_i^n$ has a slightly higher ratio of CNOT gates compared to $G_0$. However, comparing to $D_i^n$, our proposals achieve a much lower ratio of parametric gates, compensating it with superposition and two-qubit gates. Although the ratios for $D_i^n$ seem to remain constant along $n$, our approach increases the number of CNOT gates with $n$.

\begin{figure}[t]
    \centering
    \includegraphics[width=0.9\linewidth]{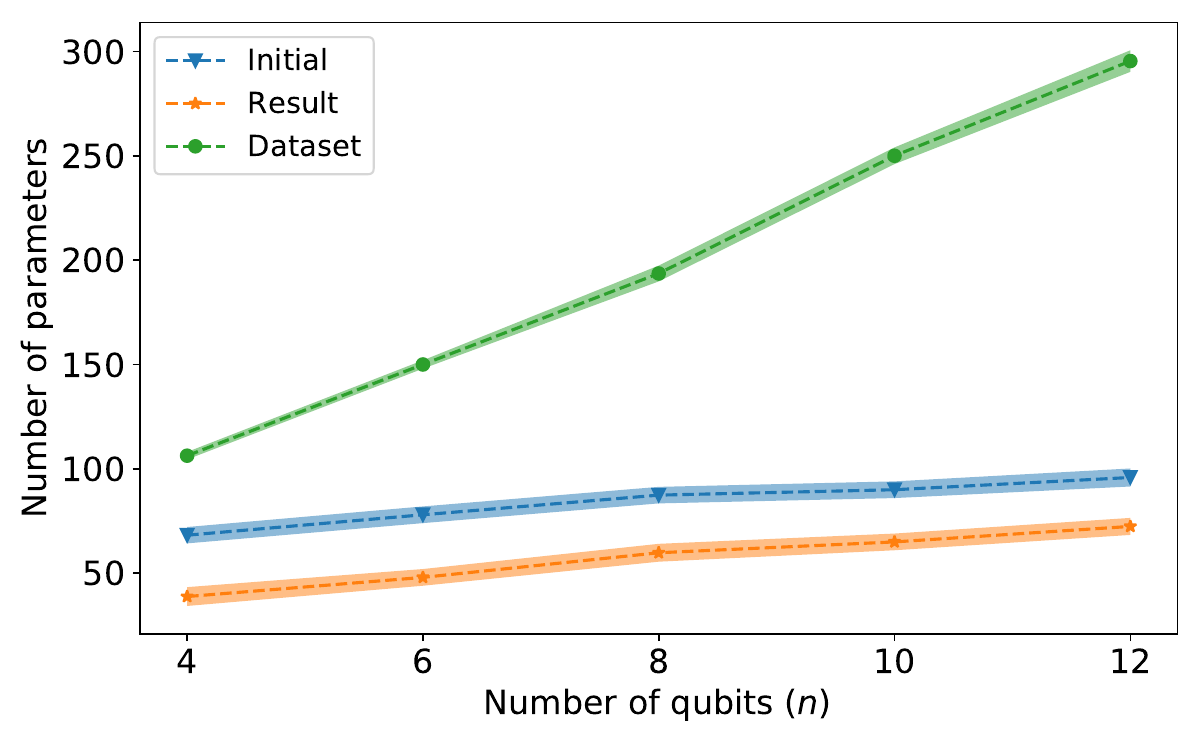}
    \caption{Mean and standard deviation of the number of parameters (Y-axis) as a function of the number of qubits (X-axis), in the \textit{ansatzes} found in the randomly generated initial data ($G_0$), best EDA solutions found ($\text{EDA}_i^n$), and dataset ($D_i^n$) \cite{nakayama2023vqegenerated}. Note that the values for $n = 6, 10$ have been approximated through a linear regression.}
    \label{fig_number_parameters_qubits}
\end{figure}

Figure~\ref{fig_number_parameters_qubits} plots the number of parameters as a function of $n$, in the set of initial randomly generated data ($G_0$), the solutions found by the EDA approach ($\text{EDA}_i^n$) and the original dataset ($D_i^n$). Although the number of gates increases linearly in the three cases, comparing the slopes found in the linear approximations of the three cases, the green function ($D_i^n$) denotes a coefficient approximately 6 times bigger than the other two functions. We show that our EDA is able to learn that a bigger number of parameters is needed, however, it does not increase this number drastically, as it is able to converge to simpler \textit{ansatz}. Shallower \textit{ansatzes} (low values in the Y-axis) are more convenient to be executed in real quantum devices due to quantum coherence and other issues of the NISQ devices.

In this experiment we tested whether our approach initialized from a random set of \textit{ansatzes} is able to converge and find similar solutions to the ones proposed in the dataset, assumed to be optimal. Figure~\ref{fig_clusters_state_vectors} and Table~\ref{tab_anova_tests} show that our algorithm finds solutions with similar state fidelity as the ones in the dataset. 

\subsection{Initialization with the dataset}
\label{sec_data_init}
The previous results have shown that the EDA approach is able to provide trainable and well performing architectures. 
In this section we initialize the EDA optimizer from the \textit{ansatzes} provided in the dataset ($D_i^n$) to test whether it is able to converge to better solutions.
Thus, the EDA execution used to face the Hamiltonian $H_i$ will be initialized using $G_0 = D_i^n$. In this case, $D_i^n$ will consist of all those architectures that meet the depth constraint imposed by the EDA. Note that, in case an architecture has a depth smaller than that imposed, the coding in binary (Equation~\ref{eq_codification}) would be equivalent to fill with identity gates (\textit{I}) until the desired depth is reached. 

The purpose of this experiment is that, given a set of \textit{ansatzes}, which are known to have good performance, we try to improve their trainability while maintaining a similar behavior. In order to compare the results found by the EDA, the energy (Equation~\ref{eq_expectation_value}) using a second level classical optimizer and the IC (Equation~\ref{eq_information_content}) are computed for all the \textit{ansatzes} in all $D_i^n$. Results are shown in Table~\ref{tab_vqe_dataset_e_ic}.

Figure~\ref{fig_pareto_frontier_qubits_label2} (Appendix) shows the Pareto frontier approximations for each $H_i$ we are facing and different numbers of $n$. 
Note that, with increasing number of qubits, the conflict between both objectives becomes more drastic. However, the EDA approach is able to identify the promising solutions in the Pareto frontier. Note that the initial generation $G_0 = D_i^n$ has been also represented to establish a reference in terms of IC. However, $\text{Score}(A)$ for the first generation should not be taken into account, as $D_i^n$ represents similar minimal energy state vectors (Figure~\ref{fig_clusters_state_vectors}), and thus, are not comparable. 

Table~\ref{tab3} (Appendix) shows the best \textit{E} and \textit{IC} found by the EDA approach where COBYLA optimizer is used, for the \textit{ansatz} parameter optimization. Note that the solutions shown in the tables are the ones that maximize $\textit{HV}$ in the Pareto frontier approximation, that is, a trade-off between both objectives in the non-dominated solutions set is found. Although in this case it is important to show the solution that optimizes the $\text{HV}$, it is possible to analyze each of the non-dominated solutions from the Pareto front in order to maximize any of the two metrics.

Regarding the results shown in Table~\ref{tab3}, it is observed a good performance in terms of expectation value minimization for $n=4$. Moreover, the IC achieved is noticeable better, which also happens in the case of $n=8$. However, the expectation value obtained for $H_3$ and $H_4$ for $n=8$ is worse than that described in the original dataset, which suggests that the EDA approach is not able to improve the metrics in Table~\ref{tab_vqe_dataset_e_ic}. 

In this experiment we tested whether our approach is able to improve the quality of the \textit{ansatz} provided in the dataset, from which the EDA is initialized. Our results show that the EDA approach is able to improve them in some of the cases, and suggest that a hyper-parameter tuning should be carried out for increasing number of qubits. 

\section{Conclusions}
\label{sec_conclusions}
In this paper we present a novel method for architecture search, in which the complexity of the multi-level optimization problem has been drastically reduced by using surrogate modelling. The EDA approach optimizes the energy estimated by the surrogate modelling by performing comparisons by pairs, and reduces the possibility of Barren plateaus issues. 

The experimental results showcase two different situations for optimizing different Hamiltonians: (i) the EDA is initialized from a random subset of solutions, and (ii) the EDA is initialized from the best solutions presented in the dataset. In the former case, the results show that the optimizer is able to converge to the same solutions presented in the dataset when the number of qubits is lower than $n=8$, and the hyper-parameters should be tuned for greater values of $n$. In the latter case, the EDA is able to improve the state of the art in some of the cases. Our approach is able to find solutions that keep a good performance regarding energy minimization, but also improve the trainability of the \textit{ansatzes} encountered. 

The numerical results analyzed suggest that the performance of our approach worsens with the number of qubits, unless the  population size ($N$) and the number of iterations ($t$) are increased. However, in order to implement a useful approach for NISQ and fault tolerant devices, the algorithm runtime for the optimization process is limited, in contrast to neural network architecture search, where the coherence of the devices do not change during time. Future work in this field would include the scalability of the algorithm to higher number of qubits ($n$).

The EDA internally uses $\text{HV}$ for ranking the architectures to be selected. Although the IC upper bound has been set based on previous experience, future work would include a dynamic definition of the reference point for the $\text{HV}$ computation, during runtime.

Given that this research is at an early stage, our primary focus is on showing underpinnings and initial feasibility rather than conducting exhaustive empirical comparisons with state-of-the-art methods. Comprehensive benchmarking and detailed empirical evaluations are planned for future studies.

\section*{Acknowledgements}
We would like to thank Yash J. Patel, Onur Danaci, Adrián Pérez-Salinas, Patrick Emonts, and the people from $\braket{aQa^L}$ group for fruitful discussions in the topic, and inviting Vicente P. Soloviev as a visitor for a few months in University of Leiden.

This work has been partially supported by the Spanish Ministry of Science and Innovation through the PID2022-139977NB-I00 project and TED2021-131310B-I00 ("Bayes-Interpret"), and by the Autonomous Community of Madrid within the ELLIS Unit Madrid framework.

This work was also partially supported by the Dutch Research Council (NWO/OCW), as part of the Quantum Software Consortium programme (project number 024.003.03), and co-funded by the European Union (ERC CoG, BeMAIQuantum, 101124342). Views and opinions expressed are however those of the author(s) only and do not necessarily reflect those of the European Union or the European Research Council. Neither the European Union nor the granting authority can be held responsible for them

Vicente P. Soloviev has been supported by the predoctoral grant FPI PRE2020-094828 from the Spanish Ministry of Science and Innovation. 

\section*{Competing interests}
The authors declare no competing interests.

\section*{Data availability}
Implementation is based on EDAspy\footnote{https://github.com/VicentePerezSoloviev/EDAspy} Python package, and the experimental scripts and data are stored in a GitHub repository\footnote{\url{https://github.com/VicentePerezSoloviev/QAS_EDA}}. The dataset used for the \textit{ansatz} comparison is published \cite{nakayama2023vqegenerated} and freely available in GitHub\footnote{\url{https://github.com/Qulacs-Osaka/VQE-generated-dataset}}.

\section*{Authorship contribution statement}
\textbf{Vicente P. Soloviev}: Conceptualization, Methodology, Software, Validation, Formal analysis, Writing – original draft. 
\textbf{Vedran Dunjko}: Project administration, Supervision, Resources, Writing – review \& editing.
\textbf{Concha Bielza}: Project administration, Supervision, Resources, Writing – review \& editing. 
\textbf{Pedro Larrañaga}: Project administration, Supervision, Resources, Writing – review \& editing.
\textbf{Hao Wang}: Project administration, Supervision, Resources, Writing – review \& editing.

\section*{Declaration of competing interest}
The authors declare that they have no known competing financial interests or personal relationships that could have appeared to influence the work reported in this paper.

\bibliographystyle{unsrt} 
\bibliography{name}

\onecolumngrid

\newpage
\appendix
\section{Hamiltonians}
\label{sec_appendix_hamiltonians}
This section describes the Hamiltonians used for the experimental results. Note that the following benchmarks and coefficients have been used in order to compare the results with the ones found in \cite{nakayama2023vqegenerated}.

1D transverse-field Ising model:
\begin{equation}
    H_1 = \sum_{i=1}^{n-1} Z_i Z_{i+1} + 2 \sum_{i=1}^{n} X_n \nonumber
\end{equation}

1D Heisenberg model:
\begin{equation}
    H_2 = \sum_{i=1}^{n-1} (X_i X_{i+1} + Y_i Y_{i+1} + Z_i Z_{i+1}) + 2 \sum_{i=1}^{n} Z_n \nonumber
\end{equation}

Su-Schrieffer-Heeger model:
\begin{equation}
    H_3 = \sum_{i=1}^{n-1} \left( 1 + \frac{3}{2} (-1)^{i-1} \right)  (X_i X_{i+1} + Y_i Y_{i+1} + Z_i Z_{i+1}) + 2 \sum_{i=1}^{n} X_n \nonumber
\end{equation}

$J_1$ - $J_2$ model:
\begin{equation}
    H_4 = \sum_{i=1}^{n-1} (X_i X_{i+1} + Y_i Y_{i+1} + Z_i Z_{i+1}) + 3 
    \sum_{i=1}^{n-2} (X_i X_{i+2} + Y_i Y_{i+2} + Z_i Z_{i+2}) \nonumber
\end{equation}

\section{Surrogate model prediction}
\label{sec_appendix_surrogate_model_prediction}
Here we compare the performance of different surrogate models by comparing different \textit{ansatzes} by pairs in a given initial data for different number of qubits. 

Different architectures have been built for problems described in Appendix~\ref{sec_appendix_hamiltonians} and different values of $n$. The number of architectures have been set to $N=37.5 n$, and the circuit depth to $m=60$.
Table~\ref{tab_accuracy_clasif_models} shows the accuracy found for different models with different configurations. Results show that support vector classifier (SVC) achieves the best metrics, and thus, is used as surrogate model in our approach.

\begin{table}[h]
    \centering
    \setlength{\tabcolsep}{10pt}
    \begin{tabular}{|c|c|c|c|}
        \hline
         model & $n=4$ & $n=8$ & $n=12$ \\
        \hline
        \hline
         Random\_forest\_20 & 0.76 & 0.77 & 0.75 \\
         Random\_forest\_50 & 0.81 & 0.82 & 0.80 \\
         Random\_forest\_80 & 0.82 & 0.83 & 0.80 \\
         KNN\_2 & 0.64 & 0.66 & 0.68 \\
         KNN\_5 & 0.72 & 0.74 & 0.75 \\
         KNN\_15 & 0.78 & 0.79 & 0.79\\
         SVC & $\bm{0.91}$ & $\bm{0.92}$ & $\bm{0.90}$ \\
         Decision tree & 0.64 & 0.65 & 0.65 \\
         Naive Bayes & 0.69 & 0.76 & 0.78 \\
        \hline
    \end{tabular}
    \caption{Accuracy found after evaluating each model in a set of initial architectures using cross-validation with 15 folds. Independently of $n$, all the \textit{ansatzes} have been restricted to $m=60$, and $N=37.5 n$. Random forest with different numbers of estimators, k-nearest neighbors (KNN) with different numbers of neighbors, support vector classifier (SVC), decision tree, and naive Bayes have been tested.}
    \label{tab_accuracy_clasif_models}
\end{table}

\newpage
\section{Distance computation}
\label{sec_appendix_distances}
Here we detail the distance comparison between all the proposed solutions within $\text{EDA}_i^n$ and each of the clusters $D_i^n$ by computing Equation~\ref{eq_distance_accuracy}. Note that index $j$ denotes each of the 5 best results found by the EDA. Table~\ref{tab_n_4_distance}-\ref{tab_n_12_distance} show the distance computations for $n \in [4, 8, 12]$, respectively.

\begin{table}[h]
    \centering
    \setlength{\tabcolsep}{12pt}
    \begin{tabular}{|c||c||c|c|c|c|}
    \hline
    \textit{ansatz} ($\text{EDA}_{i_j}^{4}$) & $H_i$ & $\text{dist}(\text{EDA}_{1_j}^{4}, D_1^{4})$ & $\text{dist}(\text{EDA}_{2_j}^{4}, D_2^{4})$ & $\text{dist}(\text{EDA}_{3_j}^{4}, D_3^{4})$ & $\text{dist}(\text{EDA}_{4_j}^{4}, D_4^{4})$ \\
    \hline
    \hline
         $\text{EDA}_{1_1}^4$ & $H_1$ & \textbf{0.018} & 0.998 & 0.990 & 0.999 \\
         $\text{EDA}_{1_2}^4$ & $H_1$ & \textbf{0.011} & 0.999 & 0.995 & 0.995 \\
         $\text{EDA}_{1_3}^4$ & $H_1$ & \textbf{0.011} & 0.999 & 0.989 & 0.999 \\
         $\text{EDA}_{1_4}^4$ & $H_1$ & \textbf{0.027} & 0.990 & 0.991 & 0.995 \\
         $\text{EDA}_{1_5}^4$ & $H_1$ & \textbf{0.011} & 0.999 & 0.989 & 0.999 \\
         $\text{EDA}_{2_1}^4$ & $H_2$ & 0.999 & \textbf{0.038} & 0.982 & 0.997 \\
         $\text{EDA}_{2_2}^4$ & $H_2$ & 0.999 & \textbf{0.049} & 0.993 & 0.999 \\
         $\text{EDA}_{2_3}^4$ & $H_2$ & 0.993 & 0.954 & \textbf{0.233} & 0.880 \\
         $\text{EDA}_{2_4}^4$ & $H_2$ & 0.999 & \textbf{0.035} & 0.976 & 0.990 \\
         $\text{EDA}_{2_5}^4$ & $H_2$ & 0.970 & \textbf{0.374} & 0.794 & 0.965 \\
         $\text{EDA}_{3_1}^4$ & $H_3$ & 0.993 & 0.999 & \textbf{0.051} & 0.660 \\
         $\text{EDA}_{3_2}^4$ & $H_3$ & 0.992 & 0.999 & \textbf{0.058} & 0.648 \\
         $\text{EDA}_{3_3}^4$ & $H_3$ & 0.988 & 0.998 & \textbf{0.064} & 0.646 \\
         $\text{EDA}_{3_4}^4$ & $H_3$ & 0.995 & 0.997 & \textbf{0.069} & 0.631 \\
         $\text{EDA}_{3_5}^4$ & $H_3$ & 0.987 & 0.999 & \textbf{0.056} & 0.637 \\
         $\text{EDA}_{4_1}^4$ & $H_4$ & 0.991 & 0.995 & 0.691 & \textbf{0.077} \\
         $\text{EDA}_{4_2}^4$ & $H_4$ & 0.993 & 0.992 & 0.752 & \textbf{0.061} \\
         $\text{EDA}_{4_3}^4$ & $H_4$ & 0.998 & 0.991 & 0.811 & \textbf{0.081} \\
         $\text{EDA}_{4_4}^4$ & $H_4$ & 0.992 & 0.997 & 0.702 & \textbf{0.099} \\
         $\text{EDA}_{4_5}^4$ & $H_4$ & 0.990 & 0.993 & 0.329 & \textbf{0.011} \\
        \hline
    \end{tabular}
    \caption{Distance (Equation~\ref{eq_distance_accuracy}) between each \textit{ansatz} in $\text{EDA}_{i_j}^4$ and $D_i^4$, where $i$ denotes the Hamiltonian index and $n=4$. Bold values represent those instances in which the closest cluster to $\text{EDA}_{i_j}^{4}$ is $D_i^4$.}
    \label{tab_n_4_distance}
\end{table}

\begin{table}[h]
    \centering
    \setlength{\tabcolsep}{12pt}
    \begin{tabular}{|c||c||c|c|c|c|}
    \hline
    \textit{ansatz} ($\text{EDA}_{i_j}^{8}$) & $H_i$ & $\text{dist}(\text{EDA}_{1_j}^{8}, D_1^{8})$ & $\text{dist}(\text{EDA}_{2_j}^{8}, D_2^{8})$ & $\text{dist}(\text{EDA}_{3_j}^{8}, D_3^{8})$ & $\text{dist}(\text{EDA}_{4_j}^{8}, D_4^{8})$ \\
    \hline
    \hline
         $\text{EDA}_{1_1}^8$ & $H_1$ & \textbf{0.973} & 0.995 & 0.995 & 0.997 \\
         $\text{EDA}_{1_2}^8$ & $H_1$ & \textbf{0.950} & 0.996 & 0.996 & 0.994 \\
         $\text{EDA}_{1_3}^8$ & $H_1$ & \textbf{0.830} & 0.998 & 0.998 & 0.998 \\
         $\text{EDA}_{1_4}^8$ & $H_1$ & \textbf{0.553} & 0.999 & 0.999 & 0.999 \\
         $\text{EDA}_{1_5}^8$ & $H_1$ & \textbf{0.942} & 0.995 & 0.990 & 0.997 \\
         $\text{EDA}_{2_1}^8$ & $H_2$ & 0.990 & \textbf{0.926} & 0.968 & 0.991 \\
         $\text{EDA}_{2_2}^8$ & $H_2$ & 0.998 & \textbf{0.906} & 0.998 & 0.999 \\
         $\text{EDA}_{2_3}^8$ & $H_2$ & 0.998 & \textbf{0.963} & 0.989 & 0.995 \\
         $\text{EDA}_{2_4}^8$ & $H_2$ & 0.996 & \textbf{0.992} & 0.998 & 0.998 \\
         $\text{EDA}_{2_5}^8$ & $H_2$ & 0.999 & \textbf{0.991} & 0.999 & 0.999 \\
         $\text{EDA}_{3_1}^8$ & $H_3$ & 0.999 & 0.999 & \textbf{0.957} & 0.995 \\
         $\text{EDA}_{3_2}^8$ & $H_3$ & 0.999 & \textbf{0.958} & 0.983 & 0.985 \\
         $\text{EDA}_{3_3}^8$ & $H_3$ & 0.999 & 0.999 & \textbf{0.522} & 0.949 \\
         $\text{EDA}_{3_4}^8$ & $H_3$ & 0.998 & 0.996 & \textbf{0.958} & 0.983 \\
         $\text{EDA}_{3_5}^8$ & $H_3$ & 0.999 & \textbf{0.922} & 0.999 & 0.996 \\
         $\text{EDA}_{4_1}^8$ & $H_4$ & 0.999 & 0.999 & \textbf{0.971} & 0.981 \\
         $\text{EDA}_{4_2}^8$ & $H_4$ & 0.999 & 0.998 & 0.992 & \textbf{0.945} \\ 
         $\text{EDA}_{4_3}^8$ & $H_4$ & 0.998 & 0.998 & \textbf{0.988} & 0.996 \\  
         $\text{EDA}_{4_4}^8$ & $H_4$ & 0.999 & 0.999 & \textbf{0.982} & 0.994 \\  
         $\text{EDA}_{4_5}^8$ & $H_4$ & 0.999 & 0.999 & 0.999 & \textbf{0.988} \\ 
         \hline
    \end{tabular}
    \caption{Distance (Equation~\ref{eq_distance_accuracy}) between each \textit{ansatz} in $\text{EDA}_{i_j}^8$ and $D_i^8$, where $i$ denotes the Hamiltonian index and $n=8$. Bold values represent those instances in which the closest cluster to $\text{EDA}_{i_j}^{5}$ is $D_i^8$.}
    \label{tab_n_8_distance}
\end{table}

\begin{table}[h]
    \centering
    \setlength{\tabcolsep}{12pt}
    \begin{tabular}{|c||c||c|c|c|c|}
    \hline
    \textit{ansatz} ($\text{EDA}_{i_j}^{12}$) & $H_i$ & $\text{dist}(\text{EDA}_{1_j}^{12}, D_1^{12})$ & $\text{dist}(\text{EDA}_{2_j}^{12}, D_2^{12})$ & $\text{dist}(\text{EDA}_{3_j}^{12}, D_3^{12})$ & $\text{dist}(\text{EDA}_{4_j}^{12}, D_4^{12})$ \\
    \hline
    \hline
         $\text{EDA}_{1_1}^{12}$ & $H_1$ & 0.999 & 0.999 & 0.999 & 0.999 \\
         $\text{EDA}_{1_2}^{12}$ & $H_1$ & 0.999 & 0.999 & 0.999 & 0.999 \\
         $\text{EDA}_{1_3}^{12}$ & $H_1$ & 0.999 & 0.999 & 0.999 & 0.999 \\
         $\text{EDA}_{1_4}^{12}$ & $H_1$ & 0.999 & 0.999 & 0.999 & 0.999 \\
         $\text{EDA}_{1_5}^{12}$ & $H_1$ & 0.999 & 0.999 & 0.999 & 0.999 \\
         $\text{EDA}_{2_1}^{12}$ & $H_2$ & 0.999 & 0.999 & 0.999 & 0.999 \\
         $\text{EDA}_{2_2}^{12}$ & $H_2$ & 0.999 & 0.999 & 0.999 & 0.999 \\
         $\text{EDA}_{2_3}^{12}$ & $H_2$ & 0.999 & 0.999 & 0.999 & 0.999 \\
         $\text{EDA}_{2_4}^{12}$ & $H_2$ & 0.999 & 0.999 & 0.999 & 0.999 \\
         $\text{EDA}_{2_5}^{12}$ & $H_2$ & 0.999 & 0.999 & 0.999 & 0.999 \\
         $\text{EDA}_{3_1}^{12}$ & $H_3$ & 0.999 & 0.998 & 0.999 & 0.999 \\
         $\text{EDA}_{3_2}^{12}$ & $H_3$ & 0.999 & 0.999 & 0.999 & 0.999 \\
         $\text{EDA}_{3_3}^{12}$ & $H_3$ & 0.999 & 0.999 & 0.999 & 0.999 \\
         $\text{EDA}_{3_4}^{12}$ & $H_3$ & 0.999 & 0.999 & 0.998 & 0.999 \\
         $\text{EDA}_{3_5}^{12}$ & $H_3$ & 0.999 & 0.999 & 0.999 & 0.999 \\
         $\text{EDA}_{4_1}^{12}$ & $H_4$ & 0.999 & 0.999 & 0.999 & 0.999 \\ 
         $\text{EDA}_{4_2}^{12}$ & $H_4$ & 0.999 & 0.999 & 0.999 & 0.999 \\      
         $\text{EDA}_{4_3}^{12}$ & $H_4$ & 0.999 & 0.999 & 0.999 & 0.999 \\      
         $\text{EDA}_{4_4}^{12}$ & $H_4$ & 0.999 & 0.999 & 0.999 & 0.999 \\      
         $\text{EDA}_{4_5}^{12}$ & $H_4$ & 0.999 & 0.999 & 0.999 & 0.999 \\     
         \hline
    \end{tabular}
    \caption{Distance (Equation~\ref{eq_distance_accuracy}) between each \textit{ansatz} in $\text{EDA}_{i_j}^{12}$ and $D_i^{12}$, where $i$ denotes the Hamiltonian index and $n=12$. }
    \label{tab_n_12_distance}
\end{table}

\section{Pareto frontier approximations}
\label{sec_appendix_pareto_frontier_qubits_label}
\begin{figure}[h]
    \centering
    \includegraphics[width=\linewidth]{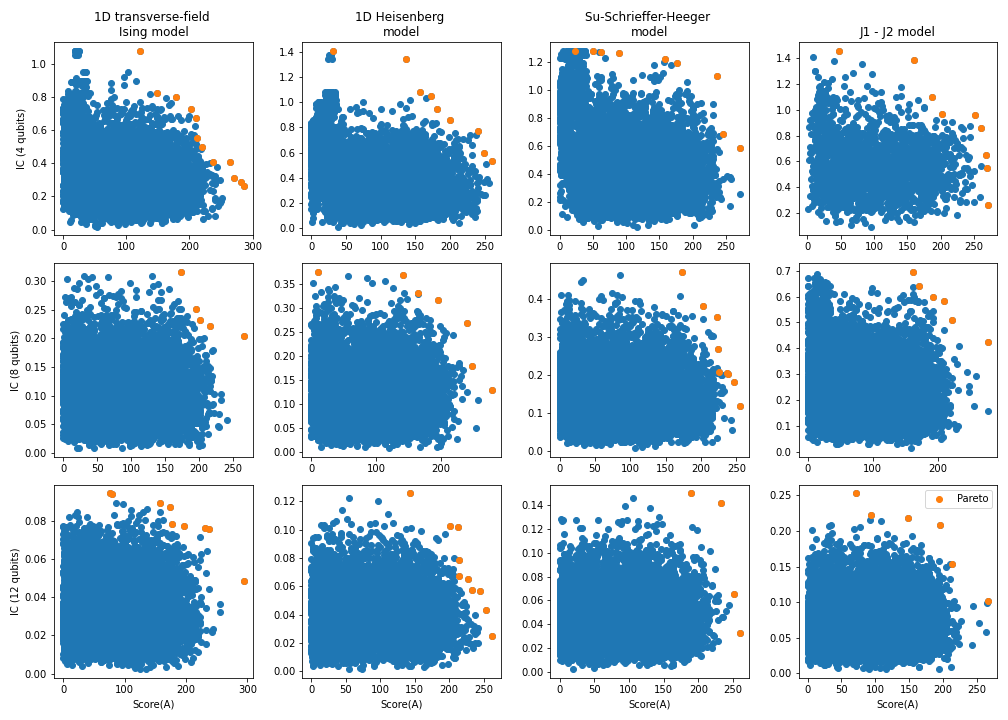}
    \caption{Pareto frontier approximation (orange spots) over all the \textit{ansatzes} considered (blue spots) during optimization process. Columns refer to problem instances, while rows refer to number of qubits ($n$).}
    \label{fig_pareto_frontier_qubits_label}
\end{figure}

Figure~\ref{fig_pareto_frontier_qubits_label} shows the Pareto frontier approximation for different $H$ and number of qubits. The columns refer to the problem instances, while the rows refer to the number of qubits ($n$). Each subplot shows all the evaluated \textit{ansatzes} (blue spots) from which the non-dominated solutions are highlighted (orange spot).

\begin{figure}[h]
    \centering
    \includegraphics[width=\linewidth]{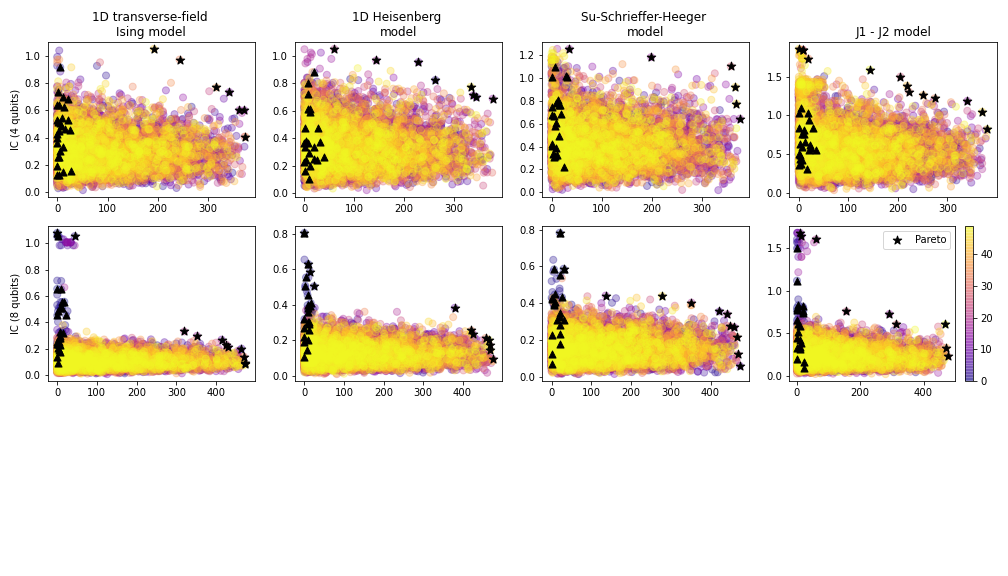}
    \caption{Pareto frontier approximation (black stars) over all the \textit{ansatzes} considered (colored spots) during the optimization process. Black triangles regard the \textit{ansatzes} included in the dataset. Columns refer to problem instances, while rows refer to number of qubits ($n$).}
    \label{fig_pareto_frontier_qubits_label2}
\end{figure}

\section{IC and expectation values comparison}
\label{appendix_e_ic_comparison}
Table~\ref{tab_vqe_dataset_e_ic} describes the mean expectation value (Equation~\ref{eq_expectation_value}) and IC (Equation~\ref{eq_information_content}) for the \textit{ansatzes} available in the dataset ($D_i^n$) for different values of $n$. 

Table~\ref{tab3} describes the best expectation value and IC found by the EDA approach for different $H_i$ and values of $n$, where the $\text{HV}$ is maximized. That is, the solutions which maximize $\text{HV}$ within $\text{EDA}_i^n$.

\begin{table}[h]
    \centering
    \setlength{\tabcolsep}{10pt}
    \resizebox{0.7\linewidth}{!}{
    \begin{tabular}{|c|c|c|c|c|}
    \hline
              & \multicolumn{2}{|c|}{$n=4$} & \multicolumn{2}{|c|}{$n=8$} \\
    \hline
              & $E$ & \text{IC} & $E$ & \text{IC} \\
    \hline
    \hline
        $H_1$ & -8.37 $\pm$ 0.01 & 0.47 $\pm$ 0.14 & -16.89 $\pm$ 0.01 & 0.46 $\pm$ 0.16 \\
        $H_2$ & -7.83 $\pm$ 0.01 & 0.51 $\pm$ 0.16 & -15.92 $\pm$ 0.02 & 0.45 $\pm$ 0.06 \\
        $H_3$ & -14.19 $\pm$ 1.87 & 0.63 $\pm$ 0.15 & -30.07 $\pm$ 0.01 & 0.51 $\pm$ 0.07 \\
        $H_4$ & -17.18 $\pm$ 2.20 & 0.80 $\pm$ 0.09 & -39.05 $\pm$ 0.04 & 0.82 $\pm$ 0.15 \\
    \hline
    \end{tabular}
    }
    \caption{ Mean and standard deviation of expectation value ($E$) (Equation~\ref{eq_expectation_value}) and information content (\text{IC}) (Equation~\ref{eq_information_content}), respectively, found in  the \textit{ansatz} in the dataset whose depth is in the range $m \pm \sqrt{m}$, for different number of qubits $n$ and Hamiltonian $H_i$. }
    \label{tab_vqe_dataset_e_ic}
\end{table}

\begin{table}[h]
    \centering
    \setlength{\tabcolsep}{10pt}
    \resizebox{0.45\linewidth}{!}{
    \begin{tabular}{|c|c|c|c|c|c|c|}
    \hline
              & \multicolumn{2}{|c|}{$n=4$} & \multicolumn{2}{|c|}{$n=8$} \\
    \hline
              & $E$ & \text{IC} & $E$ & \text{IC} \\
    \hline
    \hline
        $H_1$ & -7.81 & 0.97 & -16.18 & 0.56 \\
        $H_2$ & -6.74 & 0.73 & -13.58 & 0.45 \\
        $H_3$ & -14.03 & 1.00 & -29.28 & 0.43 \\
        $H_4$ & -17.21 & 1.47 & -26.87 & 1.57 \\
    \hline
    \end{tabular}
    }
    \caption{ Best expectation value ($E$) (Equation~\ref{eq_expectation_value}) and information content (\text{IC}) (Equation~\ref{eq_information_content}) found by the EDA approach (assisted by COBYLA) for different number of qubits ($n$) and Hamiltonians ($H_i$), where \textit{HV} is maximized in the best Pareto approximation. }
    \label{tab3}
\end{table}


\end{document}